\documentclass[float,epsfig,usenatbib]{mnras}

\usepackage{graphicx}
\usepackage{appendix}
\usepackage{psfrag}
\usepackage{amsmath,amssymb}
\usepackage{threeparttable}
\usepackage{float}
\usepackage{subfig}
\usepackage{afterpage}
\usepackage{hyperref}

\def\apj{ApJ}
\def\apjl{ApJL}
\def\mnras{MNRAS}
\def\pasp{PASP}
\def\aj{AJ}

\def\aap{A\&A}
\def\apjs{ApJS}
\def\nat{Nature}

\title[Gas-to-dust ratio across environment]{The selective effect of environment on the 
atomic and molecular gas-to-dust ratio of nearby galaxies in the {\it Herschel} Reference Survey}
\author[L. Cortese et al.]
{L. Cortese$^{1}$\thanks{luca.cortese@uwa.edu.au}, K. Bekki$^1$, A. Boselli$^2$, B. Catinella$^{1}$, L. Ciesla$^{3}$, T.~M. Hughes$^{4}$, \newauthor 
M. Baes$^{5}$, G.~J. Bendo$^{6}$, M. Boquien$^{7}$, I. de Looze$^{8}$, M.~W.~L. Smith$^{9}$, \newauthor L. Spinoglio$^{10}$, S. Viaene$^{5}$\\ 
$^1$International Centre for Radio Astronomy Research, The University of Western Australia, 35 Stirling Hwy, Crawley, WA 6009, Australia\\
$^2$Aix Marseille Universit\'{e}, CNRS, LAM (Laboratoire d'Astrophysique de Marseille) UMR 7326, 13388, Marseille, France \\
$^3$Laboratoire AIM, CEA/DSM/IRFU, CNRS, Université Paris-Diderot, 91190, Gif, France \\
$^4$Instituto de F\'{i}sica y Astronom\'{i}a, Universidad de Valpara\'{i}so, Avda. Gran Breta\~{n}a 1111, Valpara\'{i}so, Chile\\
$^5$Sterrenkundig Observatorium, Universiteit Gent, Krijgslaan 281, B-9000 Gent, Belgium\\
$^6$UK ALMA Regional Centre Node, Jodrell Bank Centre for Astrophysics, School of Physics and Astronomy,\\ The University of Manchester, Oxford Road, Manchester M13 9PL, UK\\
$^7$Unidad de Astronom\'{i}a, Fac. Cs. B\'{a}sicas, Universidad de Antofagasta, Avda. U. de Antofagasta 02800, Antofagasta, Chile\\
$^8$Department of Physics and Astronomy, University College London, Gower Street, London WC1E 6BT, UK\\
$^9$School of Physics \& Astronomy, Cardiff University, The Parade, Cardiff, CF24 3AA, UK\\
$^{10}$Istituto di Astrofisica e Planetologia Spaziali, INAF, Via del Fosso del Cavaliere 100, I-00133 Roma, Italy\\
}
\date{}
\begin{document}
\newcommand{\Zsolar}{\mbox{$\,\rm Z_{\odot}$}}
\newcommand{\Msolar}{\mbox{$\,\rm M_{\odot}$}}
\newcommand{\Lsolar}{\mbox{$\,\rm L_{\odot}$}}
\newcommand{\xs}{$\chi^{2}$}
\newcommand{\dxs}{$\Delta\chi^{2}$}
\newcommand{\xsn}{$\chi^{2}_{\nu}$}
\newcommand{\ls}{{\tiny \( \stackrel{<}{\sim}\)}}
\newcommand{\gs}{{\tiny \( \stackrel{>}{\sim}\)}}
\newcommand{\asec}{$^{\prime\prime}$}
\newcommand{\amin}{$^{\prime}$}
\newcommand{\mstar}{\mbox{$M_{*}$}}
\newcommand{\hi}{H{\sc i}}
\newcommand{\hii}{H{\sc ii}\ }
\newcommand{\kms}{km~s$^{-1}$\ }

\maketitle

\label{firstpage}

\begin{abstract}
We combine dust, atomic (\hi) and molecular (H$_{2}$) hydrogen mass measurements for 176 galaxies in the {\it Herschel} 
Reference Survey to investigate the effect of environment on the gas-to-dust mass ($M_{\rm gas}/M_{\rm dust}$) ratio of nearby galaxies. We find that, at fixed stellar mass, 
the average $M_{\rm gas}/M_{\rm dust}$ ratio varies by no more than a factor of $\sim$2 when moving from field to cluster galaxies, with Virgo galaxies 
being slightly more dust rich (per unit of gas) than isolated systems. Remarkably, once the molecular and atomic hydrogen 
phases are investigated separately, we find that \hi-deficient galaxies have at the same time 
lower $M_{\rm HI}/M_{\rm dust}$ ratio but higher $M_{\rm H_{2}}/M_{\rm dust}$ ratio than \hi-normal systems.
In other words, they are poorer in atomic but richer in molecular hydrogen if normalized to their dust content.   
By comparing our findings with the predictions of theoretical models, we show that the opposite behavior observed 
in the $M_{\rm HI}/M_{\rm dust}$ and $M_{\rm H_{2}}/M_{\rm dust}$ ratios is fully consistent with outside-in stripping of the interstellar medium (ISM), 
and is simply a consequence of the different distribution of dust, \hi\ and H$_{2}$ across the disk.
Our results demonstrate that the small environmental variations in the total $M_{\rm gas}/M_{\rm dust}$ ratio, as well as in the gas-phase 
metallicity, do not automatically imply that environmental mechanisms are not able to affect the dust and metal content of the ISM 
in galaxies. 
\end{abstract}


\section{Introduction}
The discovery, three decades ago, of a significant population of atomic hydrogen (\hi) deficient galaxies 
in nearby clusters (i.e., systems with lower \hi\ content than isolated galaxies of similar optical diameter 
and morphological type, \citealp{haynes,giova85}) provided the first clear observational 
evidence for removal of the interstellar medium (ISM) in high-density environments. 

Since then, thanks to the improvement of ground- and space-based facilities across the infrared to radio 
wavelength range, not only it has been possible to quantify more accurately the effect of the cluster 
environment on the \hi\ content of galaxies \citep{solanes01,review,cortese11}, but also we have started 
to show that environmental mechanisms can affect the other components of the ISM, such as molecular 
hydrogen (H$_{2}$), dust and metals \citep{cortese12,hughes13,boselli14env}. 
Indeed, both large statistical studies and detailed investigations of peculiar galaxies (e.g., \citealp{vollmer05,vollmer08,fumagalli09,sivanadam10,cortese10b,cortese10c,pappalardo12,pavel14,kenney15,scott15}) have shown 
that \hi-deficient systems are also, at fixed stellar mass, H$_{2}$ and dust deficient when compared to isolated galaxies. 
However, the effect on dust grains and molecules appears to be less dramatic than in the case of \hi\, supporting 
a scenario in which the ISM removal is more efficient in the outer, \hi-dominated, star-forming 
disk \citep{cortese10c,cortese12,pappalardo12,boselli14env}, as expected in the case of hydrodynamical mechanisms, such 
as ram-pressure \citep{GUNG72}.

Perhaps surprisingly, no or very little variation with environment is observed in the metal content of galaxies 
when this is quantified by means of the oxygen abundance (or gas-phase metallicity), which provides 
the amount of oxygen per unit of hydrogen in galaxies. Several works have now consistently shown that 
the variation in gas-phase metallicity (at fixed stellar mass) across environments is $\sim$0.1 dex 
at most \citep{mouchine07,ellison09,petro12,hughes13}. This observational result may suggest that the cluster 
environment is not able to affect the metal content of galaxies, in contradiction with the 
evidence supporting ram pressure stripping. 

However, this could simply come down to the fact that gas, dust and metal depletion by environment are usually 
quantified in different ways. While for gas and dust the amount of stripping is estimated 
by normalising the mass of ISM to the stellar mass, the gas-phase metallicity gives us, 
by construction, the amount of metals per unit of hydrogen mass. Thus, it is easy to see how 
the gas-phase metallicity of galaxies may not change with environment even if both metals and gas are 
removed from disk. Unfortunately, as oxygen abundance estimates do not provide an absolute quantification 
of the total amount of metals in the ISM, they cannot be easily used to test this scenario. 

The most promising way to make progress in this field is thus to investigate the effect of environment on the chemical enrichment 
of galaxies by using the gas-to-dust ratio, instead of oxygen abundance, as a proxy for the amount 
of metals per unit of hydrogen. Indeed, assuming that the dust formation and destruction timescales are similar, 
it has been shown that the gas-to-dust ratio should scale with metallicity \citep{dwek98}, as observed 
in nearby galaxies, although with significant scatter \citep{issa90,mateos09b,magrini11,remy14}. 

This technique has several advantages when compared to the oxygen abundances derived from 
optical emission lines. First, by measuring independently the total amount of gas and dust in galaxies, 
and not just their ratios as in the case of oxygen abundance, it is easier to follow the effect of 
environment on each ISM component separately. Second, while active star formation is needed to estimate gas-phase 
metallicities, far-infrared dust continuum emission relies less on the presence of star-forming regions as a significant 
fraction of dust heating comes from evolved stellar populations (e.g., \citealp{boquien11,bendo12b,bendo15}). 
Thus, we can investigate the properties of the most environmentally perturbed galaxies, which are generally 
excluded from studies of the stellar mass vs. metallicity relations simply because quiescent. 
Third, thanks to the {\it Herschel} Space Observatory we are able to trace the dust distribution across the entire 
disk, whereas gas-phase metallicity estimates have so far been generally biased towards the central regions of galaxies \citep{hughes13}.

Until very recently, this kind of analysis was limited by the lack of dust and molecular 
hydrogen estimates for representative samples spanning galaxies in different environments. 
In the last few years, we have been able to partially fill this gap by gathering this 
information for galaxies in the {\it Herschel} Reference Survey (HRS, \citealp{HRS}), a stellar mass and volume-limited sample 
of nearby systems including galaxies in the Virgo cluster. 
This provides us with the opportunity of quantifying, for the first time, the effect of the cluster 
environment on the gas-to-dust ratio of galaxies, and to try to reconcile studies of metals, dust and cold gas 
in high-density environments. 

Of course, given the limited number statistics of this sample, 
we can only compare galaxies inside and outside the Virgo cluster, and it is impossible to quantify the effect 
of environment in smaller groups and pairs. Despite this, it is important to remember that the HRS is 
currently the largest representative sample of field and cluster galaxies 
for which estimates of all phases of the ISM are available. For example, samples such 
as GASS \citep{catinella10} do not include information on the dust content of galaxies.


\begin{figure*}
\centering
\includegraphics[width=14.cm]{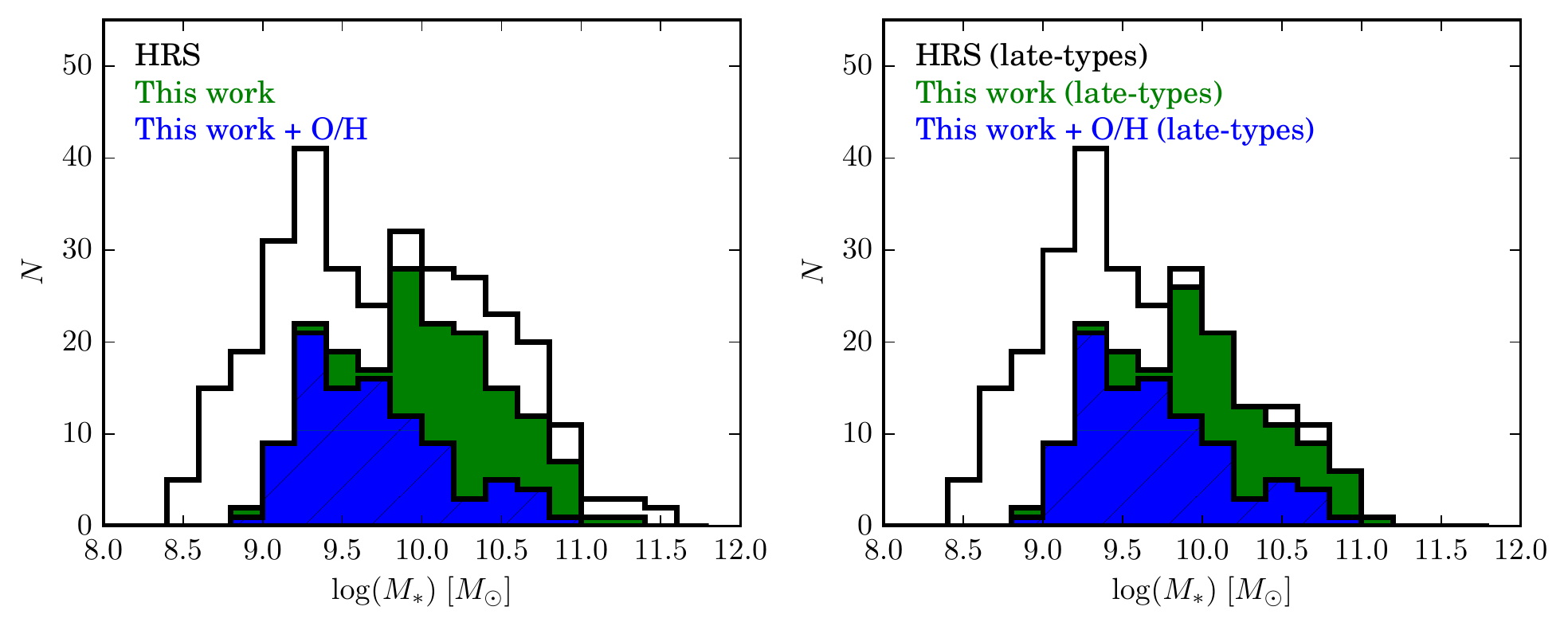}
\caption{Left: The stellar mass distribution for HRS galaxies (empty histogram) and for the sub-sample used in this work (filled green histogram). The blue hashed histogram shows the stellar mass distribution of galaxies used in this work for which also gas-phase oxygen abundance estimates are available. Right: Same as the left panel, but for late-type galaxies (Sa or later types) only. }
\label{sample}
\end{figure*}

\section{The data}
The HRS is a K-band-, volume-limited sample of 322 local galaxies. An extensive description of the 
sample selection can be found in \cite{HRS}, while updated morphologies and distance estimates are 
presented in \cite{hrsgalex}. For this work, we focus on a subsample of the HRS for which total 
cold gas (i.e., atomic plus molecular hydrogen), dust and stellar masses are available.

Atomic and molecular gas estimates are taken from \cite{hrsco}. 
\hi\ masses are derived mainly from single-dish 21 cm flux densities and are available for 315 galaxies in 
the sample: 263 detected and 52 not detected. For non-detections, 5$\sigma$ upper-limits to the \hi\ masses 
are calculated assuming a velocity width of 300 \kms. 
Molecular hydrogen masses have been estimated from CO(1-0) observations. 
Given the large uncertainty in the CO(1-0) to H$_{2}$ conversion factor ($X_{\rm CO}$), it is important 
to make sure that our results are independent of the assumptions made on $X_{\rm CO}$. Thus, 
we investigate both a constant value of $X_{\rm CO}$ throughout the sample (2.3$\times$10$^{20}$ [cm$^{-2}$ /(K km s$^{-1}$)], 
typical of Milky Way-like disks) and an H-band luminosity dependent 
conversion factor ($X_{\rm CO}$=$-$0.38 $\times$ log(L$_{\rm H}$/L$_{\odot}$) + 24.23 [cm$^{-2}$ /(K km s$^{-1}$)]) as calibrated in \cite{boselligdust}.
The H-band is mainly used here as a proxy for metallicity in order to be able to follow the effect of a metallicity-dependent $X_{\rm CO}$ without 
the need of oxygen abundance estimates for all our galaxies.  
Out of the 225 galaxies observed in CO(1-0), 143 have been detected. For the remaining 82 objects, 
5$\sigma$ upper limits have been determined by assuming a CO velocity width equal to the \hi\ width, 
when available, or to 300 \kms in case of \hi\ non detections. 

Dust masses have been estimated as follows. For the 262 HRS galaxies detected across the entire 8 $\mu$m to 500 $\mu$m wavelength range, 
we use the values presented in \cite{ciesla2014}, which have been obtained by fitting the \cite{draine07b} dust models to the infrared spectral 
energy distribution. For the 38 galaxies not detected in neither of the three {\it Herschel}/SPIRE bands, we estimate 5$\sigma$ upper-limits 
to their dust masses following the method presented in \cite{cortese12}. These are based on the 350$\mu$m flux densities assuming a dust emissivity 
coefficient ($\beta$) equal to 2. The remaining 22 galaxies, for which a complete dust spectral energy distribution cannot be determined, 
are excluded from our sample to avoid possible biases in the dust mass estimates. 
Although the use of the \cite{draine07b} dust models requires a wider wavelength coverage (8-500 $\mu$m) than what needed 
to fit a single modified black-body to the {\it Herschel} SPIRE and PACS bands (100-500 $\mu$m), we note that this requirement does not introduce 
any significant bias to our sample. Indeed, out of the 22 galaxies detected in at least one {\it Herschel} band but excluded from our sample, 
cold gas mass estimates are available for 15 objects, and for only 6 of these a modified black-body fit with $\beta$=2 (needed to be consistent 
with the \citealp{draine07b} dust models) is successful in providing a dust mass estimate (see \citealp{cortese14} for a description of single modified black-body 
fits to HRS galaxies).

Finally, stellar mass estimates are taken from \cite{hrsgalex}, which were obtained from the Sloan Digital Sky Survey $g$ and $i$ band magnitudes 
using the empirical recipes presented in \cite{zibetti09}.

In order to look for any variations in the $M_{\rm gas}/M_{\rm dust}$ ratio of galaxies across environment, we focus our attention on the subsample 
of HRS galaxies for which stellar, dust, atomic and molecular gas mass estimates are available, and for which at least one of the three ISM components 
has been detected. This second condition is crucial for deriving meaningful upper (or lower) limits to the $M_{\rm gas}/M_{\rm dust}$ ratio. 
Thus, our sample is composed of 176 HRS galaxies. For 122 objects ($\sim$70\% of our final sample) all three ISM phases 
have been detected, for 41 galaxies we have detections of both dust and one gas component (35 in \hi\ and 6 in H$_{2}$ - we will 
refer to these galaxies as `interval censored data'), whereas for the remaining 13 objects we either detect only 
dust (5 galaxies) or only gas (8 systems). 
Thus the $M_{\rm gas}/M_{\rm dust}$ ratio remains poorly constrained for only $\sim$7\% of our sample, as for interval censored 
data we are able to determine both upper and lower limits to the real value.

Our final sample is approximately half of the entire HRS, and it is important to determine whether it is still representative of the entire survey. 
We investigate this in Fig.~\ref{sample}, where we compare the stellar mass distribution 
for the entire HRS (empty histogram) with that of the sample used in this paper (filled green histogram). For stellar masses 
greater than $\sim$10$^{9.2}$ M$_{\odot}$, our completeness is 70\% or higher across all stellar masses, whereas it rapidly decreases for smaller galaxies. 
This low completeness for dwarf galaxies is mainly due to the lack of CO(1-0) observations \citep{hrsco}. 
Moreover, our completeness further improves at high stellar masses, if we consider late-type galaxies only (i.e., Sa or later-types; right panel in Fig.~\ref{sample}).
A simple Kolmogorov-Smirnov (KS) test suggests that we cannot reject the null hypothesis that 
the stellar mass distribution of our sample is drawn from the one of the 
full HRS above a stellar mass of $\sim$10$^{9}$ M$_{\odot}$ at a $\sim$25\% level, suggesting that our selection criteria have not introduced significant biases.  

In Fig.~\ref{sample}, we also show the stellar mass distribution of galaxies in our final sample for which gas-phase oxygen abundances on the \cite{pettini04} base metallicity calibration can be determined from emission lines (blue hashed histogram; \citealp{hughes13}). The large difference between the blue and the green histograms (96 versus 176 galaxies) highlights the benefits of using the $M_{\rm gas}/M_{\rm dust}$ ratio, instead of gas-phase metallicity, to investigate the enrichment history of galaxies, particularly those with stellar mass greater than $\sim$10$^{10}$ M$_{\odot}$. Indeed, in this case the KS test rejects the null hypothesis at a $\sim$0.5\% level.

\begin{figure*}
\centering
\includegraphics[width=15.cm]{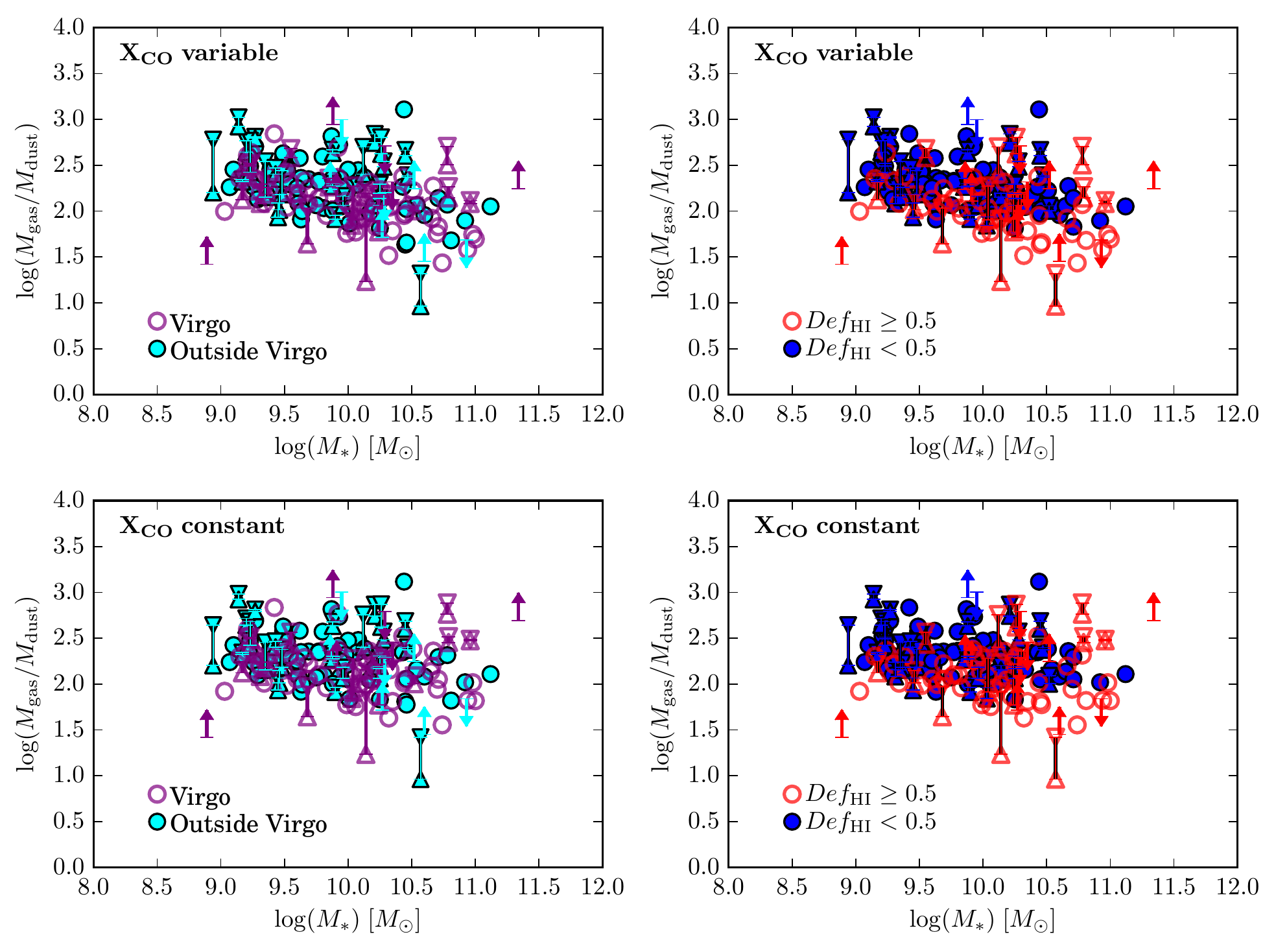}
\caption{The $M_{\rm gas}/M_{\rm dust}-M_{*}$ relation for galaxies in the HRS. Circles and arrows indicate 
detections and upper-limits, respectively. Interval censored data (i.e., galaxies for which dust and only one gas phase are detected) 
are presented as connected triangles, showing the possible range of variation in $M_{\rm gas}/M_{\rm dust}$ obtained by setting 
the non detected phase to either zero or its 5$\sigma$ upper limit. 
The left column shows a comparison between Virgo cluster members (empty purple) and galaxies outside the Virgo cluster (filled cyan), 
whereas in the right column galaxies are divided according 
to their \hi\ content, with \hi-deficient and \hi-normal galaxies as red empty and blue filled symbols, respectively. The top and bottom rows 
show results obtained for a variable and constant CO-to-H$_{2}$ ($X_{\rm CO}$) conversion factor, respectively.}
\label{gd}
\end{figure*}

\section{The total gas-to-dust ratio across environment}
We define the total $M_{\rm gas}/M_{\rm dust}$ ratio as 1.3$\times$$(M_{\rm HI}+M_{\rm H_{2}})/M_{\rm dust}$, where the 1.3 coefficient is included to account 
for the contribution of Helium. Only if, out of the three ISM components, just one of the two gas phases has been 
detected, conservative lower limits to the gas-to-dust ratio are defined as 1.3 $\times$ $M_{\rm HI}/M_{\rm dust}$ and 1.3$\times$$M_{\rm H_{2}}/M_{\rm dust}$, 
for \hi-only or H$_{2}$-only detections, respectively. 

Throughout this paper we will refer to the gas-to-dust ratio as $M_{\rm gas}/M_{\rm dust}$, instead of GDR as typically done in the 
literature. This is to remind the reader that here we investigate total mass ratios, whereas the original definition 
of GDR (in both observational and theoretical works) is based on the ratio of column densities. Since the distribution 
of the three ISM components in galaxies is not the same (see also Sec.~4), the total mass and surface density 
ratios may differ. 

Following \cite{cortese12}, we investigate the effect of environment in two complementary ways. 
First, we divide galaxies according to their Virgo cluster membership (Fig.~\ref{gd}, left panels): i.e., Virgo cluster members (84 galaxies) 
and galaxies outside Virgo (92 galaxies). Second, we use the \hi-deficiency parameter ($Def_{\rm HI}$) as an environmental 
indicator (Fig.~\ref{gd}, right panels). Def$_{\rm HI}$ is defined as the logarithmic difference between the expected \hi\ mass for an isolated galaxy with the same morphological type and optical diameter of the target and the observed value \citep{haynes}, and it is generally adopted to 
identify environmentally perturbed systems. We use a value of Def$_{\rm HI}$=0.5 (i.e., 70\% less \hi\ than expected) to discriminate between \hi-stripped galaxies (69 galaxies) and `unperturbed' systems (107 galaxies)\footnote{It is important to remember that, although the vast majority of \hi-deficient galaxies (54 galaxies) 
are either Virgo member or lie in its infalling regions, some of them are found in groups (see also \citealp{catinella13}). However, our conclusions are unchanged if we remove from our sample \hi-deficient galaxies outside Virgo.}. The difference between the top and bottom rows is simply due to the different CO to H$_{2}$ conversion factor used: 
$X_{\rm CO}$ varying with H-band luminosity (top), or $X_{\rm CO}$ constant (bottom).
\begin{figure*}
\centering
\includegraphics[width=15.cm]{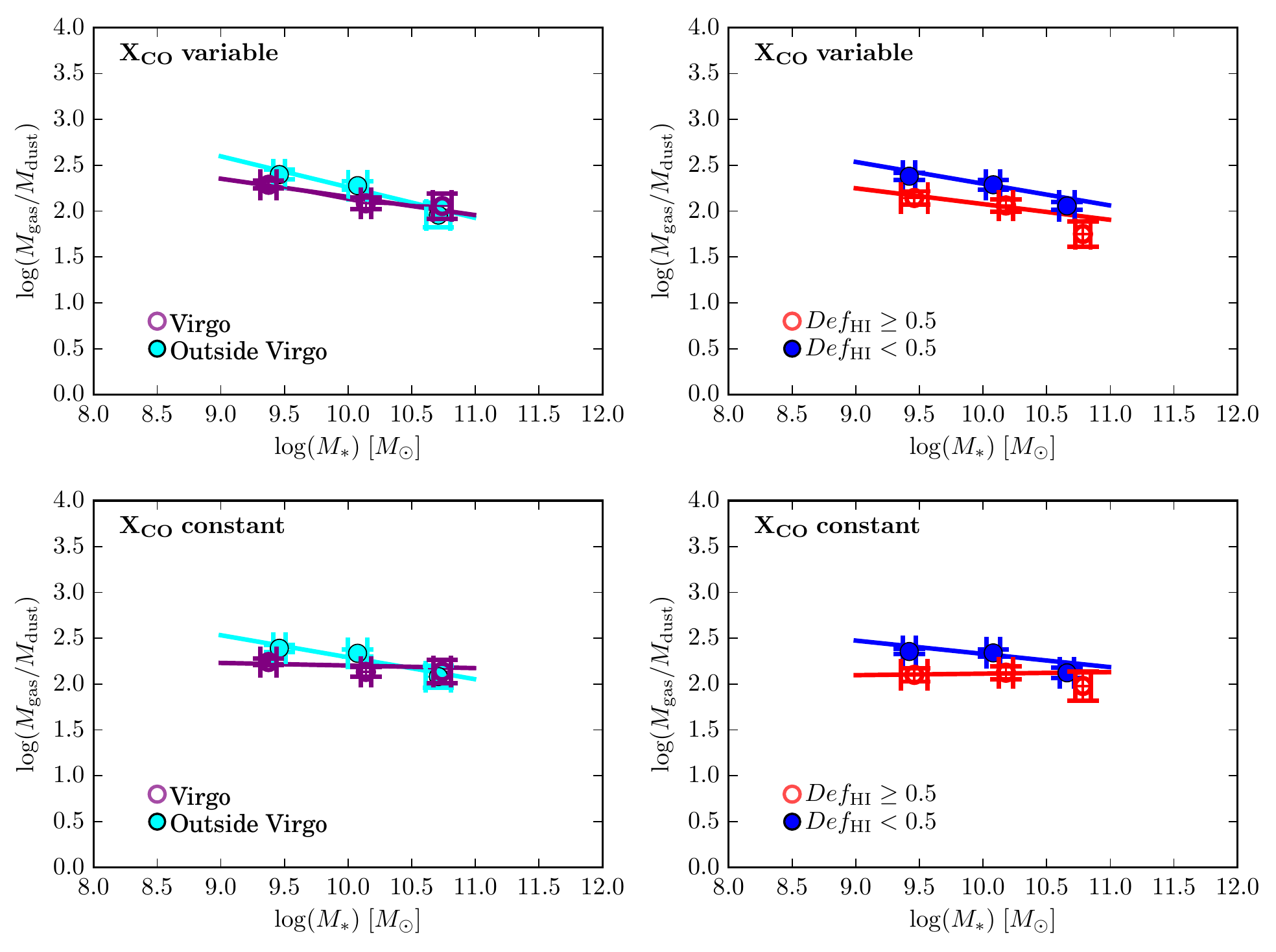}
\caption{The $M_{\rm gas}/M_{\rm dust}-M_{*}$ relation for galaxies in the HRS. Points show median values per bin of stellar mass, whereas lines show the best linear fits to the data. Errors on the median are obtained using 1000 bootstrap re-samplings of the data in each bin (see Table~1). Panels and colours are as in Fig.~\ref{gd}.}
\label{gd_fits}
\end{figure*}

\begin{table*}
\caption {The median $M_{\rm gas}/M_{\rm dust}$ ratio per bin of M$_{*}$ for HRS galaxies outside Virgo, Virgo cluster members, \hi-normal and \hi-deficient systems. 
Errors on the median are obtained using 1000 bootstrap re-samplings of the data in each bin.}
\[
\label{gdtab}
\begin{array}{cccc}
\hline\hline
\noalign{\smallskip}
\noalign{\smallskip}
\log(M_{*}/M_{\odot})^{a} &   \log(M_{\rm gas}/M_{\rm dust})  & \log(M_{\rm gas}/M_{\rm dust})  & N_{\rm gal}   \\                  
              &    X_{\rm CO}~constant & X_{\rm CO}~variable &    \\
\noalign{\smallskip}
\hline
\multicolumn{4}{c}{\rm Outside~Virgo}\\
\noalign{\smallskip}
  9.46	     &  2.39\pm0.04   & 2.40\pm0.06      &   37	  \\
  10.08	     &  2.34\pm0.04   & 2.28\pm0.04      &   44	  \\
  10.71	     &  2.08\pm0.12   & 1.96\pm0.14      &   11	  \\
\noalign{\smallskip}
\hline
\multicolumn{4}{c}{\rm Virgo~members}\\
\noalign{\smallskip}
  9.38	     &  2.24\pm0.04   & 2.29\pm0.04      &   32	  \\
  10.14	     &  2.13\pm0.05   & 2.09\pm0.06      &   37	  \\
  10.74	     &  2.14\pm0.13   & 2.06\pm0.14      &   15	  \\
\noalign{\smallskip}
\hline
\multicolumn{4}{c}{{\rm HI-normal}~(Def_{\rm HI}<0.5)}\\
\noalign{\smallskip}
  9.42	     &  2.36\pm0.03     	& 2.38\pm0.04     &   52	  \\
  10.08	     &  2.34\pm0.04     	& 2.29\pm0.05     &   45	  \\
  10.66	     &  2.12\pm0.06     	& 2.06\pm0.04     &   10	  \\
\noalign{\smallskip}
\hline
\multicolumn{4}{c}{{\rm HI-deficient}~(Def_{\rm HI}\geq0.5)}\\
\noalign{\smallskip}
  9.46	     & 2.10\pm0.08     	& 2.14\pm0.08     &   17	  \\
  10.18	     & 2.12\pm0.07     	& 2.06\pm0.07     &   36	  \\
  10.79	     & 1.98\pm0.16     	& 1.75\pm0.14     &   16	  \\
\hline
\hline
\end{array}
\]
$^{a}$ Stellar mass bins are identical for each subsample: i.e., $log(M_{*}/M_{\odot})<9.8$, $9.8\leq log(M_{*}/M_{\odot})\leq 10.5$ and $log(M_{*}/M_{\odot})>10.5$ .
\end{table*}

\begin{figure*}
\centering
\includegraphics[width=15.cm]{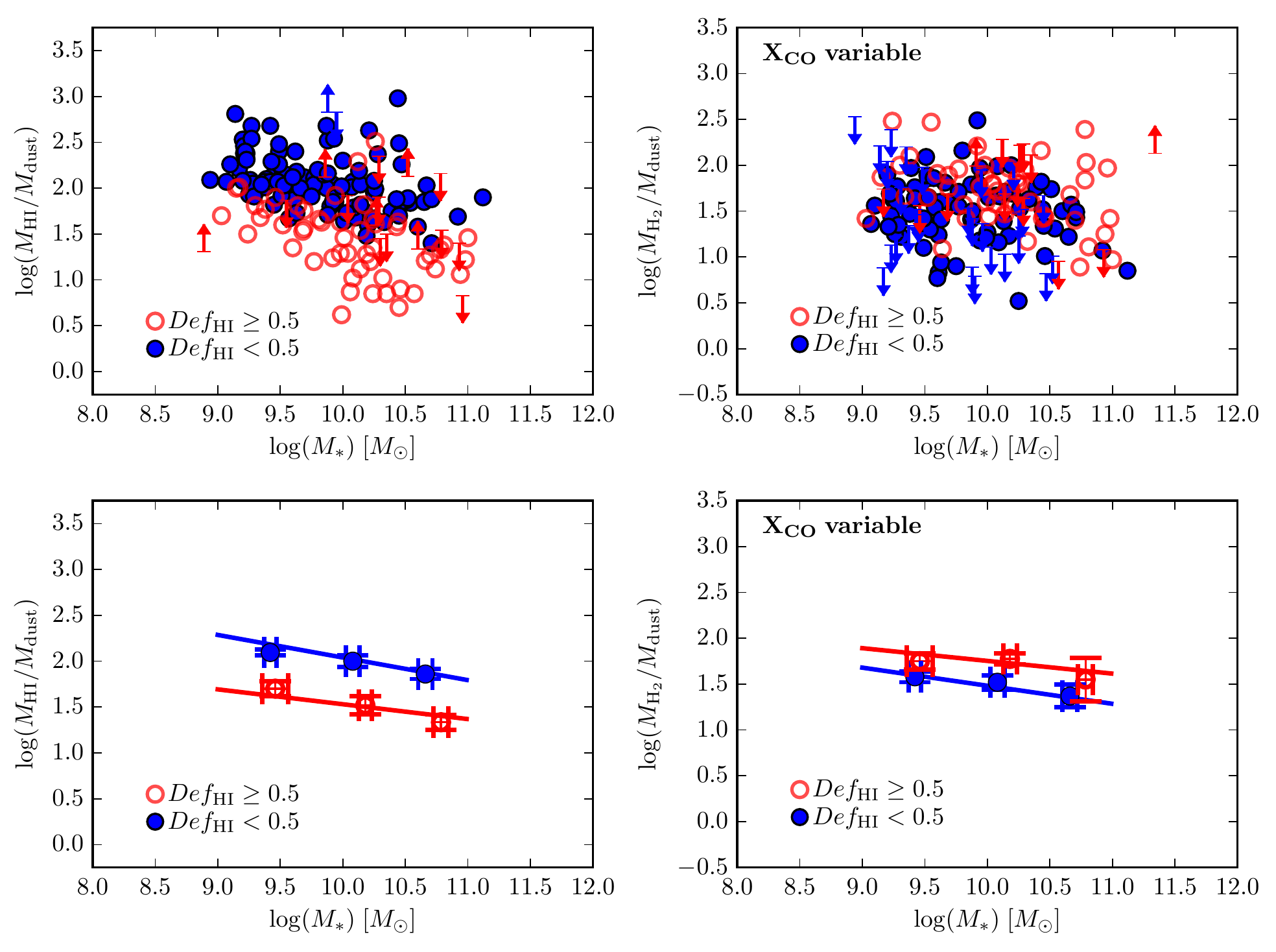}
\caption{The $M_{\rm HI}/M_{\rm dust}$-$M_{*}$ (left) and $M_{\rm H_{2}}/M_{\rm dust}$-$M_{*}$ (right) relations for \hi-normal (filled blue) and 
\hi-deficient (empty red) HRS galaxies. Individual galaxies are shown in the upper panels, with circles and arrows 
indicating detections and upper/lower limits, respectively. Averages per bin of stellar mass and best linear fits and 
shown in the bottom panels.}
\label{dhi_dh2}
\end{figure*}

In all panels, circles show the 122 galaxies for which all the three ISM phases are detected, and arrows indicate the 13 objects for which either only 
gas or dust are detected. The remaining 41 galaxies, for which dust and one gas phase are detected, are presented 
as connected triangles, where the two triangles correspond to the upper and lower limit to the $M_{\rm gas}/M_{\rm dust}$ obtained 
by setting the undetected gas phase to its 5$\sigma$ upper limit or to zero, respectively.

As recently shown by \cite{remy14}, we confirm that the $M_{\rm gas}/M_{\rm dust}$ ratio monotonically decreases with stellar mass regardless of environment. 
The slope of the relation depends on the value of X$_{\rm CO}$ used but, even for a constant conversion factor, we 
see an average variation in the $M_{\rm gas}/M_{\rm dust}$ ratio of $\sim$0.2 dex when moving from less ($\sim$10$^{9.4}$ M$_{\odot}$) to more 
massive ($\sim$10$^{10.7}$ M$_{\odot}$) systems, with a scatter of 0.25-0.3 dex.

\begin{table*}
\caption {The median $M_{\rm HI}/M_{\rm dust}$-M$_{*}$ and $M_{\rm H_{2}}/M_{\rm dust}$-M$_{*}$ relations for \hi-normal and \hi-deficient galaxies. 
Errors on the median are obtained using 1000 bootstrap re-samplings of the data in each bin.}
\[
\label{hih2tab}
\begin{array}{ccccc}
\hline\hline
\noalign{\smallskip}
\noalign{\smallskip}
\log(M_{*}/M_{\odot}) &  \log(M_{\rm HI}/M_{\rm dust}) &  \log(M_{\rm H_{2}}/M_{\rm dust})  & \log(M_{\rm H_{2}}/M_{\rm dust})  & N_{\rm gal}   \\                  
              &                   &    X_{\rm CO}~constant & X_{\rm CO}~variable &    \\
\noalign{\smallskip}
\hline
\multicolumn{5}{c}{{\rm HI-normal}~(Def_{\rm HI}<0.5)}\\
\noalign{\smallskip}
  9.42	     & 2.10\pm0.03    &  1.47\pm0.09     & 1.58\pm0.06         &   52	  \\
  10.08	     & 2.00\pm0.06    &  1.73\pm0.10     & 1.52\pm0.08         &   45	  \\
  10.66	     & 1.86\pm0.05    &  1.68\pm0.11     & 1.37\pm0.13         &   10	  \\
\noalign{\smallskip}
\hline
\multicolumn{5}{c}{{\rm HI-deficient}~(Def_{\rm HI}\geq0.5)}\\
\noalign{\smallskip}
  9.46	     & 1.70\pm0.08  &         1.70\pm0.11   & 1.75\pm0.09      &   17	  \\
  10.18	     & 1.52\pm0.10  &         1.84\pm0.09   & 1.78\pm0.06      &   36	  \\
  10.79	     & 1.33\pm0.07  &         1.91\pm0.22   & 1.55\pm0.25      &   16	  \\
\hline
\hline
\end{array}
\]
\end{table*}

In order to quantify the effect of environment on the $M_{\rm gas}/M_{\rm dust}$-$M_{*}$ relation 
in Fig.~\ref{gd_fits} we present the best-linear fits for each subsample as well 
as the median and their bootstrapped errors per bin of stellar mass. 
We fix interval censored data to their upper limits and left/right censored data to their upper/lower-limits, 
but our conclusions are independent on how interval censored data are treated.

Regardless the choice of X$_{\rm CO}$, 
when galaxies are divided according to their cluster membership, an effect of the environment is evident only 
in the lower stellar mass bin, with cluster galaxies having $\sim$0.15 dex lower $M_{\rm gas}/M_{\rm dust}$ ratio than field systems. Indeed, at higher 
stellar masses, the difference in the average $M_{\rm gas}/M_{\rm dust}$ ratio for the two samples is within the errors.
This is in line with previous results based on the mass-metallicity relations of cluster galaxies (e.g., \citealp{mouchine07,ellison09,petro12,hughes13}), and in particular with the recent work by \cite{petro12}, who found that the effect of the cluster environment on the gas-phase metallicity becomes significant only for $M_{*}<$10$^{9.5}$ M$_{\odot}$. 
When galaxies are separated according to their \hi\ content, the effect of environment becomes slightly more significant ($\sim$2-2.5 $\sigma$ level on average) 
with \hi-deficient galaxies having a $M_{\rm gas}/M_{\rm dust}$ ratio $\sim$0.2-0.3 dex lower than \hi-normal systems, 
at all stellar masses, independently of the choice of $X_{\rm CO}$.

These results show that the effect of the environment on the $M_{\rm gas}/M_{\rm dust}$ is rather limited and becomes evident only if we 
pre-select galaxies that have already been affected by the environment based on their \hi\ deficiency measurement. 
This is in line with the lack of strong environmental 
dependence observed in the stellar mass vs. gas-phase metallicity relation \citep{hughes13}. As extensively discussed 
in \cite{cortese12}, this is simply due to the fact that cluster membership does not imply that environment has already 
been able to play a significant role.

In order to further investigate why the $M_{\rm gas}/M_{\rm dust}$ ratio shows at most a factor of two variation with environment 
despite the fact that the \hi\ reservoir of \hi-deficient galaxies has been depleted by more than 
a factor of three, we now consider the gas-to-dust ratio for the \hi\ and H$_{2}$ phases, separately. 
In Fig.~\ref{dhi_dh2}, which represents the main result of this work, 
we plot the $M_{\rm HI}/M_{\rm dust}$ (left) and $M_{\rm H_{2}}/M_{\rm dust}$ (right) ratios as 
a function of stellar mass for our sample. Individual galaxies are shown in the top row, while 
median per bin of stellar mass and best linear fits are presented in the bottom row. 
We only focus on the comparison between \hi-deficient (red) and \hi-normal (blue) galaxies, 
as this is the case for which the effect of environment is stronger. We also show the case of 
variable $X_{\rm CO}$ factor only, but the results for a constant value of $X_{\rm CO}$ are included in Table \ref{hih2tab}.

The difference between \hi-poor and \hi-normal galaxies is larger 
once the atomic and molecular components are considered separately. 
In particular, the M$_{\rm HI}$/M$_{\rm dust}$ ratio is $\sim$0.5 dex lower in \hi-deficient galaxies (see Table \ref{hih2tab}). 
As already discussed in \cite{cortese12}, this is consistent with the idea that the \hi, 
extending to larger radii than te dust, is more strongly affected by the environment. 
The improvement with respect to our previous work is simply in the use of more accurate \hi\ and dust mass 
estimates. Remarkably, the $M_{\rm H_{2}}/M_{\rm dust}$ ratio shows the {\it opposite trend} (see Fig.~\ref{dhi_dh2} left), with 
\hi-deficient systems having systematically higher $M_{\rm H_{2}}/M_{\rm dust}$ ratios than \hi-normal galaxies at fixed stellar 
mass ($\sim$0.2 dex). Moreover, both the $M_{\rm HI}/M_{\rm dust}-M_{*}$ and $M_{\rm H_{2}}/M_{\rm dust}$-$M_{*}$ relations 
have a scatter slightly larger ($\sim$0.35 dex) than the $M_{\rm gas}/M_{\rm dust}-M_{*}$ trend ($\sim$0.25-0.3 dex).
As shown in Table~\ref{hih2tab}, the slope of the $M_{\rm H_{2}}/M_{\rm dust}$-$M_{*}$ depends on the choice of 
$X_{\rm CO}$ (i.e., decreasing/increasing with stellar mass for $X_{CO}$ variable/constant), as expected if the 
$M_{\rm H_{2}}/M_{\rm dust}$ ratio is roughly constant with metallicity. 
Conversely, the difference between \hi-poor and \hi-normal systems does not depend on $X_{\rm CO}$. 

Admittedly, while the change in the M$_{\rm HI}$/M$_{\rm dust}$ ratio has a high statistical significance, it 
becomes marginal (i.e., $\sim$1-1.5$\sigma$ level on average) for the $M_{\rm H_{2}}/M_{\rm dust}$ ratio. This is due to the larger 
error in the CO measurements and aperture corrections, as well as the higher fraction of non-detections.

We can safely exclude that our findings are an artifact due to the aperture correction applied 
to the CO data (see \citealp{hrsco}). Indeed, the typical fractional coverage of the CO 
observations is roughly the same for both classes of galaxies, and the same result is 
obtained if we consider the uncorrected data. 

By combining the results presented in the two panels of Fig.~\ref{dhi_dh2},  
it is clear why the $M_{\rm gas}/M_{\rm dust}$ ratio varies so little with environment: the reduction  
in M$_{\rm HI}$/M$_{\rm dust}$ ratio seems partly balanced by an increase in M$_{\rm H2}$/M$_{\rm dust}$ for cluster galaxies. 
In other words, the absence of a large variation in the $M_{\rm gas}/M_{\rm dust}$ ratio (and similarly of gas-phase metallicity) 
as function of environment does not automatically imply that the environment is not able to 
affect the different phases of the ISM. On the contrary, each phase is removed 
in such a way that the net effect on the $M_{\rm gas}/M_{\rm dust}$ ratio 
is relatively small.

\section{Differential stripping of the ISM}
A natural scenario that could explain the results presented in Fig.~\ref{dhi_dh2} is stripping 
by ram-pressure. The hydrodynamic pressure of the intra-cluster medium on galaxies 
removes primarily the ISM in the outer-parts of the disk, which are generally \hi-dominated compared 
to the inner effective radius where most of the neutral hydrogen is condensed into molecules. 
Thus, if environment is more efficient in the outskirts of galaxies, the net effect on the total gas 
and dust reservoirs is simply related to how extended their distributions are. 
In other words, in case of ram pressure, the evidence for differential stripping shown 
in the previous section may be simply a consequence of the different scale-lengths of the 
\hi , H$_{2}$ and dust disks. 

We test this hypothesis in two different, complementary, ways. 
In the first one, we use an analytical approximation to show how we can reproduce 
our findings with simple assumptions on the distribution of gas and dust in the disk. 
In the second one, we take advantage of the \cite{bekki13c,bekki14,bekki14b} model to follow the effect of ram pressure 
on the various components of the ISM in a self-consistent way.

\begin{figure*}
\centering
\includegraphics[width=17.5cm]{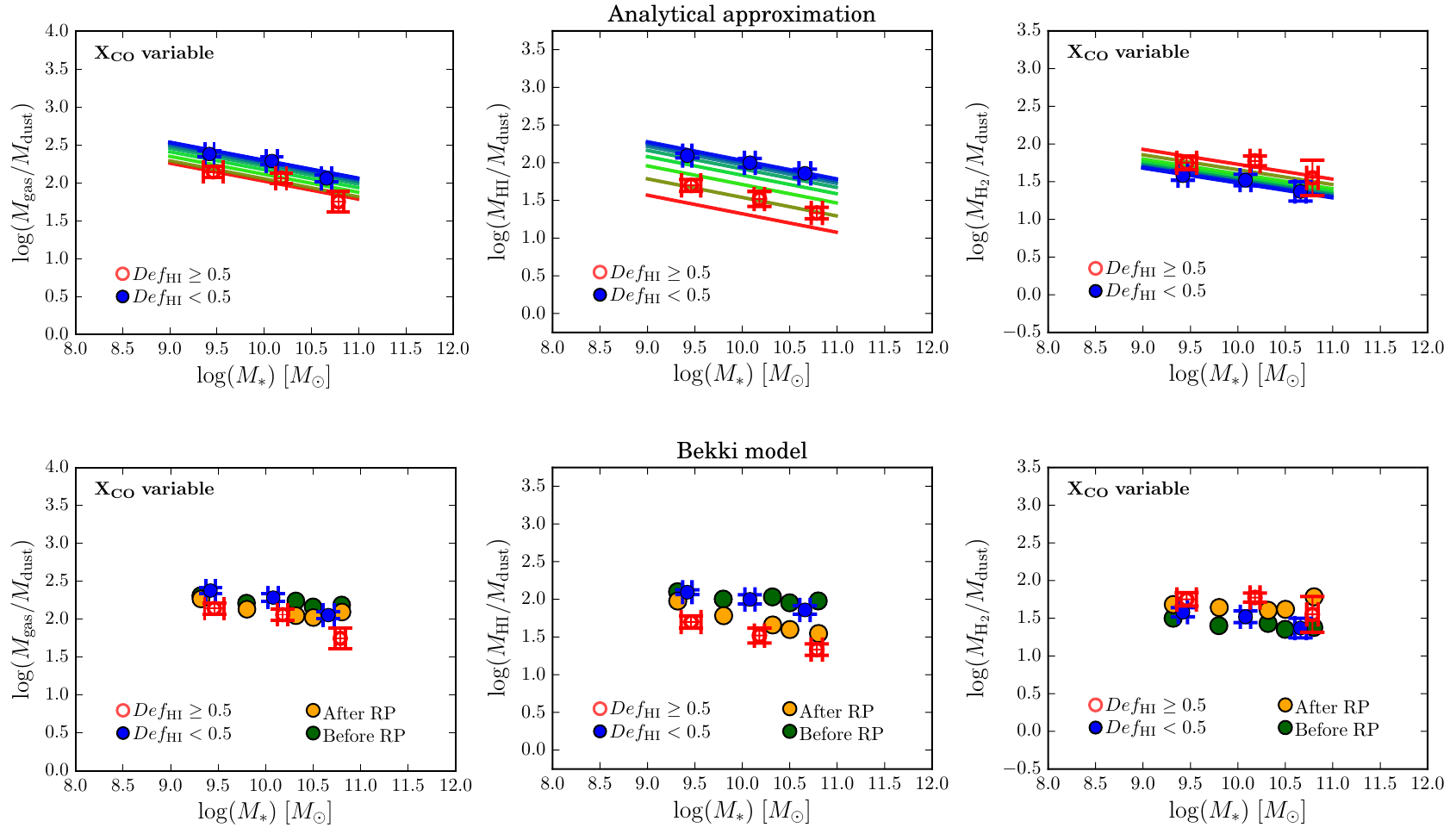}
\caption{The average $M_{\rm gas}/M_{\rm dust}$ (left), $M_{\rm HI}/M_{\rm dust}$ (centre) and $M_{\rm H_{2}}/M_{\rm dust}$ (right) ratios per bin of M$_{*}$ for 
\hi-normal (filled blue) and \hi-deficient (empty red) galaxies, reproduced from Fig.~3 and 4. The top panel shows the predictions for our simple analytical model 
where all ISM components are removed outside the stripping radius. Coloured lines show stripping radii decreasing 
from 2$\times R_{\rm opt}$ (i.e., no stripping, blue) to 0.3$\times R_{\rm opt}$ (red), in 0.2$\times R_{\rm opt}$ steps. 
The bottom panels show the predictions of Bekki (2013, 2014a, b) models. Green and yellow circles show the predictions 
before and after ram-pressure stripping.}
\label{models}
\end{figure*}

\subsection{Analytical approximation}
The basic assumption behind our analytical approach is that all three components 
of the ISM are fully stripped by ram pressure up to the same stripping radius. 
Thus, to determine how a galaxy would move in Figs.~\ref{gd} and \ref{dhi_dh2}, we 
simply need to estimate how much mass is left within the stripping radius. 

For the \hi\ distribution we assume the functional form presented in \cite{hewitt83} 
by studying a sample of fifty-two nearby galaxies mapped with the Arecibo telescope:  
\begin{equation}
\Sigma_{\rm HI}(r) = 3 e^{\frac{-r^{2} } {R_{0}^{2}}} - 1.8 e^{\frac{-r^{2}}{0.23 R_{0}^{2}}}
\end{equation}
and fix $R_{0}$ equal to the optical radius ($R_{\rm opt}$). 
The double Gaussian shape is used to reproduce the central depression in the \hi\ surface density 
profile, where most of the gas has already condensed into molecules, as typically 
observed in late-type galaxies \citep{leroy08}. 

Conversely, we consider an exponential surface density distribution for 
both dust and H$_{2}$
\begin{equation}
\Sigma(r) = \Sigma(0) e^{-r/R_{d}}
\end{equation}
with $R_{d}(dust)$=R$_{\rm opt}$/3.2 (i.e., the dust follows the stellar mass distribution, \citealp{mateos09b}; Smith et al., submitted) 
and  $R_{d}(H_{2})$=0.2R$_{\rm opt}$ \citep{schruba11}.

In the ideal case where all the material outside the stripping radius ($R_{\rm strip}$)
is removed by ram-pressure, we can determine the reduction in the mass of each component 
after stripping as 
\begin{equation}
\frac{M_{\rm after}}{M_{\rm before}} = \frac{\int_{0}^{R_{\rm strip}} 2\pi r \Sigma(r) dr}{ \int_{0}^{\infty} 2\pi r \Sigma(r) dr }
\end{equation}
For the atomic hydrogen this gives 
\begin{equation}
\label{eqhi}
\frac{M_{\rm after}}{M_{\rm before}} =  1 - 1.16e^{-\frac{R_{\rm strip}^2}{R_{0}^2}}  + 0.16e^{-\frac{R_{\rm strip}^2}{0.23R_{0}^2}} 
\end{equation}
whereas for dust and H$_{2}$
\begin{equation}
\label{eqdust}
\frac{M_{\rm after}}{M_{\rm before}} = 1 - (1 + \frac{R_{\rm strip}}{R_{d}})  e^{-\frac{R_{\rm strip}}{R_{d}}} 
\end{equation}
Taking advantage of Eqs. \ref{eqhi} and \ref{eqdust}, we can now determine how a galaxy will shift in Figs.~\ref{gd} and \ref{dhi_dh2}, 
depending on the size of the stripping radius. We assume the best-fitting linear relation for \hi-normal galaxies 
as the reference in case of unperturbed systems and consider values of $R_{\rm strip}$ in the range 0.3-2 $R_{\rm opt}$ (0.2 wide steps).
We also consider an average molecular-to-atomic fraction of $\sim$27\%, i.e., the average value for our \hi-normal galaxies.

The expected shifts in the $M_{\rm gas}/M_{\rm dust}$, $M_{\rm HI}/M_{\rm dust}$ and $M_{\rm H_{2}}/M_{\rm dust}$ ratios for different stripping radii are presented 
in the top panel of Fig.~\ref{models}. It emerges that, under the simple assumptions discussed above, we can quantitatively 
reproduce the difference between \hi-deficient and \hi-normal galaxies in $M_{\rm gas}/M_{\rm dust}$, $M_{\rm HI}/M_{\rm dust}$ and $M_{\rm H_{2}}/M_{\rm dust}$ ratios 
simultaneously, by assuming a typical stripping radius for \hi-deficient galaxies of $\sim$0.5 R$_{\rm opt}$.
For the \hi\ component, this degree of stripping is in line with what is observed in \hi-deficient galaxies in Virgo. Indeed, by taking advantage of 
the data presented in \cite{chung09} (see also Fig.~2 in \citealp{cortese10c}), it is easy to show that $Def_{\rm HI}>$0.5 
implies $R_{\rm strip}<$0.9 $R_{\rm opt}$. Interestingly, this simple model is also able to quantitatively reproduce the correlation 
between \hi\ and H$_{2}$ deficiency observed by \cite{boselli14env}, with H$_{2}$ deficiencies above $\sim$0.3 only when 
$Def_{\rm HI}>$1.

\subsection{Bekki's model}
Although physically motivated, the analytical approximation described above is rather empirical and not self-consistent. 
In particular, it does not take into account the possible small increase of star formation due to 
gas compression by ram pressure, and it naively assumes that ram-pressure equally affects all three ISM components 
outside the stripping radius, which may not always be true \citep{pappalardo12}. 
Thus, in this section we compare our findings with the predictions of the model developed by \cite{bekki13c,bekki14,bekki14b}. 
Although this model is not set in a cosmological framework, it is the only model currently available where the evolution 
of the three different phases, and the effect of environment, is followed in a self-consistent way. 
Cosmological simulations are only now starting to include dust formation models \citep{mckinnon15}, as so far they have 
just assumed a linear correlation between gas-to-dust ratio and metallicity. 

This model allows us to investigate  spatial and temporal variations of gas and 
dust components (carbon and silicate dust)  in disk galaxies
with different masses and Hubble types. The code adopts the smoothed-particle hydrodynamics 
method to follow the time evolution of gas dynamics in galaxies.

An extensive description of the model is provided in \cite{bekki13c,bekki14,bekki14b}, here 
we briefly summarise its main features. 
A disk galaxy is modeled as a fully self-gravitating
system composed of  dark matter halo, stellar and gaseous disks,
and stellar bulge. 
Although we investigated numerous models, we show only five representative 
models with dark matter halos of 10, 5, 3.3, 1 and 0.33$\times$10$^{11}$ M$_{\odot}$  
and disk stellar masses of 5.4, 2.7, 1.8, 0.54 and 0.18$\times$10$^{10}$ M$_{\odot}$. 
We will explore other models, and their implications
on dust and gas evolution in cluster environments in a future paper.

The gas disk is assumed to have an exponential
profile with its size initially twice the stellar disk
and the gas mass fraction is set to be 0.18.
The initial gaseous metallicities of disk galaxies are calibrated 
on the mass-metallicity relation, spanning the range -0.11$<[Fe/H]<$0.34, 
with a negative radial metallicity gradient of $-0.04$ dex kpc$^{-1}$.

Dust formation from ejecta of AGB stars and supernovae is self-consistently implemented. 
The initial dust-to-metal ratio is set to
be 0.4 (as observed for the Milky Way), and the initial $M_{\rm gas}/M_{\rm dust}$ ratio 
gradient follows the metallicity gradient.
In order to avoid further introduction of free parameters on dust,
we do not include the dust growth and destruction for $\sim 1$ Gyr
evolution of disk galaxies under ram pressure of intra-cluster medium in the present study.

The mass of cold gas present in the molecular phase is determined by 
the balance between H$_{2}$ condensation on dust grains from atomic gas 
and photo-dissociation by the far-ultraviolet radiation field (see Eq.~16 in \citealp{bekki13c}). 
For a given gas density, total dust-to-gas-ratio, and intensity of the interstellar 
radiation field, the H$_{2}$ mass fraction associated to each particles, and 
thus the galaxy H$_{2}$ profile, can be determined.

The strength of ram pressure force on the disk 
is time-dependent and modeled in the same way as in \cite{bekki14}.
We consider a Virgo-like cluster with a total mass of 
10$^{14}$ ${\rm M}_{\odot}$ and a temperature of the ICM $T_{\rm ICM}=2.6 \times 10^7$ K, 
in line with the observed values for the Virgo A cloud \citep{boeringer1994,schindler99}.
The total ICM mass is 15\% of the dark matter mass of the cluster and both dark matter 
and ICM are assumed to follow a \cite{nfw} profile.
In order to show the effects of ram pressure on the ISM we choose the 
orbits of disk galaxies for which stripping can be efficient, but part of the 
gas is left within the disk after the first passage (otherwise no $M_{\rm gas}/M_{\rm dust}$ ratio can be estimated). 
The initial velocity of the galaxy is 617 \kms, but this increases to 1246 \kms 
at the pericentre, which is set to 30 kpc from the cluster centre.

%

The predictions of this model are shown in the bottom row of Fig.~\ref{models}. Green and yellow circles 
illustrate the results before and after (1 Gyr from the pericentre) ram-pressure stripping. Even once the balance between 
\hi, H$_{2}$ and dust, as well as the effect of the environment, are modeled 
in a self-consistent way we reproduce the opposite trends observed in the $M_{\rm HI}/M_{\rm dust}$ and 
$M_{\rm H_{2}}/M_{\rm dust}$ ratios. While the quantitative agreement is quite good for the $M_{\rm gas}/M_{\rm dust}$ and 
$M_{\rm H_{2}}/M_{\rm dust}$ ratios, the model seems to slightly underestimate the stripping of \hi\ 
especially in low mass galaxies. However, this is likely simply due to the particular choice of orbits 
assumed here. We stress that the goal of this exercise is to demonstrate that the 
differential stripping scenario is consistent with our data, not to determine which set of model 
parameters best fits our observations. There are many assumptions and free variables in the modeling 
that could be tweaked in order to perfectly match our observations, but looking for 
an exact fit is beyond the goal of this work. 

To conclude, both the theoretical approaches presented here confirm that the changes in the 
$M_{\rm gas}/M_{\rm dust}$ ratios observed when moving from \hi-normal to \hi-deficient galaxies 
are consistent with simple outside-in stripping of the ISM. 

Of course, at this stage we cannot exclude that more than one environmental mechanism is responsible 
for the trends presented in this paper (as a small fraction of our \hi-deficient galaxies are in groups). However, 
all the observational evidence collected so far suggests that stripping by a hydrodynamical mechanism 
is the main factor to explain the bulk of our \hi-deficient population.


\begin{figure}
\centering
\includegraphics[width=7.2cm]{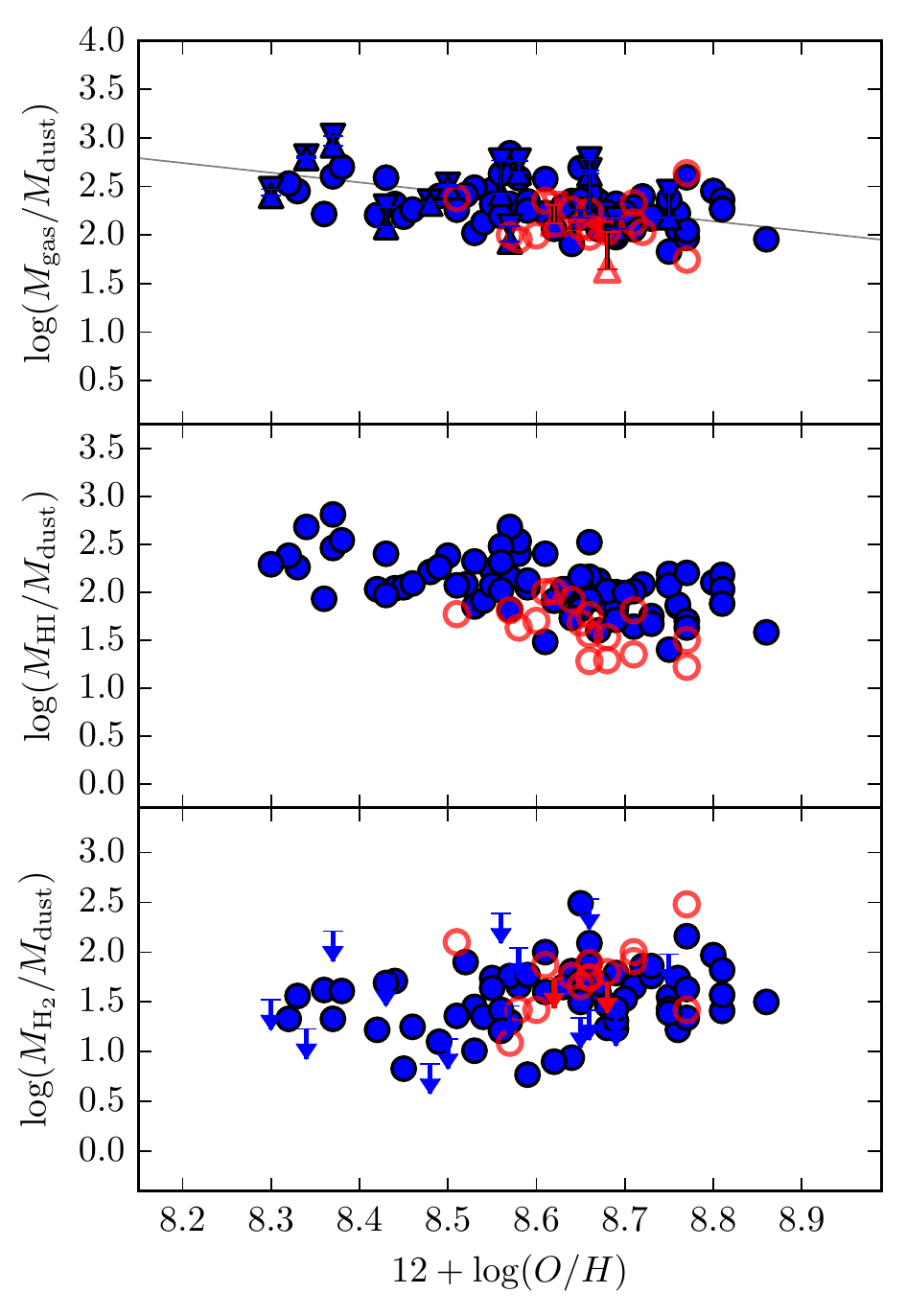}
\caption{The $M_{\rm gas}/M_{\rm dust}$ (top), $M_{\rm HI}/M_{\rm dust}$ (middle) and $M_{\rm H_{2}}/M_{\rm dust}$ ratios a function of gas-phase metallicities 
on the PP04 O3N2 base calibration. Symbols are as in the right column of Fig.~2. Note that this plot includes only the subsample of galaxies for which oxygen abundances 
are available (blue histogram in Fig.~1). The grey solid line in the top panel shows the case $M_{\rm gas}/M_{\rm dust}\propto$ O/H.}
\label{gd_oh}
\end{figure}

\section{Summary \& Conclusion}
In this work we have performed the largest investigation on the effect of environment on the $M_{\rm gas}/M_{\rm dust}$ ratio of nearby galaxies.
We have provided evidence for significant reduction in the dust, atomic and molecular hydrogen reservoirs of 
cluster galaxies, once compared with field systems of the same stellar mass. This ISM removal results in a systematic decrease in 
the $M_{\rm HI}/M_{\rm dust}$ ratio and increase in the $M_{\rm H_{2}}/M_{\rm dust}$ ratio for gas-poor systems. However, once both atomic 
and molecular hydrogen are combined, the net effect of the environment is reduced and, at fixed stellar mass, 
the $M_{\rm gas}/M_{\rm dust}$ ratio varies by $\sim$0.2 dex at most from low- to high-density environments. 

Our results do reconcile the clear evidence for gas and dust stripping in clusters, 
with the absence of any strong environmental trends observed in the stellar mass vs. metallicity relation. 
Indeed, given the tight link between $M_{\rm gas}/M_{\rm dust}$ ratio and gas-phase metallicity 
(e.g., \citealp{burnstein82,issa90,remy14}) observed also in our sample (Fig.~\ref{gd_oh}, top panel), it is natural to see 
how the tiny change in the oxygen abundance of galaxies (at fixed stellar mass) with environment 
does not automatically imply that the environment has no effect on the metal content of galaxies. 
This is particularly true when galaxies are simply divided according to their cluster membership.  
On the contrary, both metals and hydrogen are affected, but in such a way that their ratio remains nearly constant. 

We have shown that the observed variations of $M_{\rm gas}/M_{\rm dust}$, $M_{\rm HI}/M_{\rm dust}$ and $M_{\rm H_{2}}/M_{\rm dust}$ ratios 
with environment are consistent with what expected if a hydrodynamical mechanism such as ram-pressure stripping is the main 
responsible for the removal of the ISM from the disk. 
In particular, we have highlighted how our findings can be simply understood as a consequence 
of the different spatial distribution of atomic, molecular hydrogen and dust across a galaxy: i.e., 
with molecular hydrogen being more centrally concentrated than dust, and dust more concentrated than 
\hi. Although several previous works have found evidence of differential stripping in cluster galaxies 
(e.g., \citealp{kenney86,vollmer09,tonnesen10,cortese10c,cortese12,sperello13,boselli14env,kenney15}), this is the first time that the strengths 
of this effect is quantified for a large statistical 
sample and for all three phases of the ISM simultaneously. 
This work, combined with our previous studies of the HRS \citep{hughes09,cortese09,cortese11,cortese12,hughes13,boselli14env}, allows us to complete the 
analysis of the effect of Virgo cluster environment on the different steps of the star formation cycle 
of galaxies, confirming that direct stripping of ISM from the disk (e.g., ram pressure) is the main 
mechanism affecting the star formation cycle of cluster galaxies at the present epoch (see also \citealp{boselli14guvics}). 

Finally, it is worth noting the possible implications of our results for the use of dust continuum emission 
as a proxy for cold gas content \citep{eales2010,eales12,groves15}, a technique that is becoming more and more common 
in high redshift studies \citep{magdis12,scoville14}. The small, but significant (i.e., comparable to the scatter of the relation) 
variation of the $M_{\rm gas}/M_{\rm dust}$ ratio with environment might suggest that the total dust mass does not always represent an 
ideal proxy for the total cold hydrogen content of environmentally perturbed galaxies. 
More importantly, the larger differences between \hi-deficient and \hi-normal galaxies observed 
for the $M_{\rm HI}/M_{\rm dust}$ and $M_{\rm H_{2}}/M_{\rm dust}$ ratios (as well as the larger intrinsic scatter in the relations) 
shows that, at fixed stellar mass, dust emission cannot be easily used to quantify the amount of hydrogen 
present in either molecular or atomic phase, as the three phases of the ISM are distributed 
differently across the disk. This is particularly important considering the fact that, when moving 
from $z=0$ to $z\sim$1, the \hi/H$_{2}$ ratio is expected to decrease significantly, the gas 
reservoirs in galaxies at $z\sim$1 are thought to be mainly in the molecular phase \citep{lagos11,popping14}, 
and the physical conditions, as well as distribution, of the ISM are remarkably different from those of nearby disks.

However, the above argument does not take into account that the calibration of the dust continuum emission 
as a gas tracer is generally based on the $M_{\rm gas}/M_{\rm dust}$-$O/H$ relation and not on the $M_{\rm gas}/M_{\rm dust}$-$M_{*}$ relation. 
Since stripping is able to affect the amount of metals in the ISM, but should not change significantly the stellar mass of galaxies, 
it is possible that the effect of the environment on a $M_{\rm gas}/M_{\rm dust}$-$O/H$ diagram would be less strong. 
Unfortunately, as discussed in \S~2, for nearly half of our sample gas-phase 
metallicities cannot be estimated, making this kind of test unreliable. Despite this, for completeness we show in Fig.~\ref{gd_oh} the 
$M_{\rm gas}/M_{\rm dust}$ (top), $M_{\rm HI}/M_{\rm dust}$ (middle) and $M_{\rm H_{2}}/M_{\rm dust}$ (bottom) ratios as a function of gas-phase metallicities (taken 
from \citealp{hughes13}) for \hi-normal and \hi-deficient galaxies. Given the limited number of gas-poor galaxies available 
in this sample we cannot draw firm conclusions, but it is interesting to note that a difference in the $M_{\rm H_{2}}/M_{\rm dust}$ and 
$M_{\rm HI}/M_{\rm dust}$ of \hi-normal and \hi-deficient galaxies is still present. Moreover, it appears that the anti-correlation between 
$M_{\rm gas}/M_{\rm dust}$ and $O/H$ is primarily driven by the atomic gas phase, whereas very little dependence on metallicity is 
observed in case of the $M_{\rm H_{2}}/M_{\rm dust}$. 

We can thus conclude that, until the relation between gas-to-dust ratio and environment is directly 
investigated in high-redshift galaxies, it remains unclear whether or not the relations between dust and cold gas 
calibrated on local galaxies provide us with an unbiased view of the gas cycle in galaxies at higher redshifts.

\section*{Acknowledgments}
We thank the referee for a constructive report which helped improving the quality of this manuscript.

LC and BC acknowledge support under the Australian Research Council’s Discovery Programme funding schemes 
(FT120100660, DP130100664, DP150101734). M. Boquien acknowledges funding by the FIC-R Fund,  allocated to
the project 30321072.

LC thanks V. Lebouteiller and J. Fernandez-Ontiveros for comments on this manuscript. 

We thank all the people involved in the construction and the launch of {\it Herschel}.
PACS has been developed by a consortium of institutes led by MPE (Germany) and including UVIE (Austria); KU Leuven, CSL, IMEC (Belgium); 
CEA, LAM (France); MPIA (Germany); INAF-IFSI/OAA/OAP/OAT, LENS, SISSA (Italy); IAC (Spain). This development has been supported 
by the funding agencies BMVIT (Austria), ESA-PRODEX (Belgium), CEA/CNES (France), DLR (Germany), ASI/INAF (Italy), 
and CICYT/MCYT (Spain). SPIRE has been developed by a consortium of institutes led by Cardiff University (UK) and including Univ. Lethbridge (Canada); NAOC (China); CEA, LAM (France); IFSI, Univ. Padua (Italy); IAC (Spain); Stockholm Observatory (Sweden); Imperial College London, RAL, UCL-MSSL, UKATC, Univ. Sussex (UK); and Caltech, JPL, NHSC, Univ. Colorado (USA). This development has been supported by national funding agencies: CSA (Canada); NAOC (China); CEA, CNES, CNRS (France); ASI (Italy); MCINN (Spain); SNSB (Sweden); STFC (UK); and NASA (USA)

\end{document}